\documentclass[12pt]{article}

\usepackage{latexsym}
\usepackage{amssymb}
\usepackage{amsbsy}
\usepackage{amsmath}
\usepackage{epsfig}
\usepackage{graphics,color}
\usepackage{a4}
\usepackage{cite}    
\numberwithin{equation}{section}

\newcommand{\be}{\begin{equation}}
\newcommand{\ee}{\end{equation}}
\newcommand{\bea}{\begin{eqnarray}}
\newcommand{\eea}{\end{eqnarray}}
\newcommand{\bean}{\begin{eqnarray*}}
\newcommand{\eean}{\end{eqnarray*}}
\newcommand{\nn}{\nonumber}

\newcommand{\bra}{\langle}
\newcommand{\ket}{\rangle}
\newcommand{\eps}{\epsilon}

\newcommand{\Tr}{\mbox{Tr}\,}
\newcommand{\Trr}{\mbox{Tr}}

\newcommand{\tht}{\phi}
\newcommand{\thtR}{\phi^\rmR}
\newcommand{\thtI}{\phi^\rmI}
\newcommand{\re}{\mbox{Re}\,}
\newcommand{\im}{\mbox{Im}\,}

\newcommand{\rmR}{{\rm R}}
\newcommand{\rmI}{{\rm I}}

\newcommand{\half}{\frac{1}{2}}
\newcommand{\om}{\omega}

\begin{document}

\title{ \bf\large
 Complex Langevin dynamics in the SU(3) spin model at nonzero chemical 
potential revisited
}

\author{
\addtocounter{footnote}{2}
 Gert Aarts\thanks{email: g.aarts@swan.ac.uk} \;
  and 
 Frank A.\ James\thanks{email: pyfj@swan.ac.uk} \\
\addtocounter{footnote}{1}
\mbox{} \\
 {\em\normalsize Department of Physics, College of Science, Swansea University} \\
 {\em\normalsize Swansea, United Kingdom} \\
}

\date{December 20, 2011}
\maketitle

\begin{abstract}
The three-dimensional SU(3) spin model is an effective Polyakov loop model for QCD at nonzero temperature and density. 
It suffers from a sign problem at nonzero chemical potential.
We revisit this model using complex Langevin dynamics and assess in particular 
the justification of this approach, using analyticity at small $\mu^2$ 
 and the criteria for correctness developed recently.
Finite-stepsize effects are discussed in some detail and a higher-order algorithm is employed to 
eliminate leading stepsize corrections. 
Our results strongly indicate that complex Langevin dynamics is reliable in this theory
in both phases, including the critical region. This is in sharp contrast to the case of the XY model, 
where correct results were obtained in only part of the phase diagram.

\end{abstract}

\maketitle

%%%%%%%%%%%%%%%%%%%%%%%%%%%%%%%%%%%%%%%%%%%%%%%%%%%

\newpage

\section{Introduction}
\label{sec:intro}

The phase structure of QCD as temperature and baryon chemical potential are varied has not yet been determined from first 
principles \cite{arXiv:1111.5370}.
Due to the presence of the sign problem (at nonzero chemical potential the fermion determinant is complex), the cornerstone of numerical lattice gauge theory, importance sampling, is not applicable.
In the past decade several approaches have been put forward to access the phase diagram at small quark chemical potentials $\mu$ and at temperatures near the transition temperature  between the confined and the quark-gluon plasma phase. While in agreement when $\mu \lesssim T$, 
none of these methods can be extended to larger $\mu$ values (see Refs.\ \cite{pdf,arXiv:1101.0109} for recent reviews).
Given that current and upcoming experiments at the Relativistic Heavy Ion Collider at BNL, the Large Hadron Collider at CERN and the Facility for Antiproton and Ion Research at GSI aim to map out the phase boundaries in the QCD phase diagram by colliding heavy ions at relativistic speeds, there is ample motivation to study the sign problem and ways to resolve it, both in QCD and in related theories.

There are several methods which allow the sign problem to be eliminated altogether, but these are not universally applicable.
In some theories it is possible to group degrees of freedom together in such a way that the sign problem is manifestly absent. This is the  idea behind the meron cluster algorithm \cite{arXiv:cond-mat/9902128}  and it has been applied recently to random matrix theory at finite chemical potential \cite{arXiv:1103.3467}. 
A constrained sampling of field space, yielding a joint probability distribution for only a small number of observables, is the notion behind the factorization/density of states/histogram approaches
\cite{hep-lat/0208025,Fodor:2007vv,arXiv:0706.3549,arXiv:1009.4504}.
 Sometimes it is possible to reformulate a theory in terms of  dual variables in a sign-problem free manner 
\cite{Chandrasekharan:2008gp,arXiv:1001.3648}. Recent successful applications of this have been to models derived from QCD in combined strong-coupling and heavy-quark expansions 
\cite{arXiv:0907.1915,arXiv:1010.0951,arXiv:1102.3096,arXiv:1104.2503,arXiv:1111.0916,arXiv:1111.1434,arXiv:1111.4953}.
 
In this paper we consider complex Langevin dynamics, a numerical algorithm not relying on importance sampling but
instead on a complexification of the degrees of freedom, resulting in new ways to explore field space \cite{Parisi:1984cs,Klauder:1983,Damgaard:1987rr}.
 We consider the three-dimensional SU(3) spin model at nonzero density, an effective Polyakov loop model which follows from the QCD Lagrangian in a combined strong-coupling and heavy-quark expansion and one of the first QCD-related models addressed with complex Langevin  dynamics \cite{KW,BGS}.
Our reason to revisit this model is partly due to the recent discussion of Gattringer, who showed how a reformulation in terms of  fluxes eliminates the sign problem \cite{arXiv:1104.2503}.
Moreover, given that our understanding of complex Langevin dynamics has steadily improved in the past years
\cite{Berges:2005yt,Berges:2006xc,Berges:2007nr,Aarts:2008rr,Aarts:2008wh,Aarts:2009hn,arXiv:0912.0617,arXiv:0912.3360,arXiv:1005.3468,arXiv:1006.0332,arXiv:1101.3270}, 
we consider it worthwhile to reconsider the model and apply recently developed tools \cite{arXiv:0912.3360,arXiv:1005.3468,arXiv:1101.3270}
to assess the applicability of complex Langevin dynamics in detail, something that was not undertaken in the classic papers \cite{KW,BGS}.

This paper is organized as follows. In the next section we introduce the SU(3) model and summarize some basic results at finite density. The complex Langevin equations are given in Sec.\ \ref{sec:dcl}. Besides the standard lowest-order discretization, we also describe a higher-order algorithm to eliminate leading stepsize corrections \cite{CCC}.
 In Sec.\ \ref{sec:cfc} we discuss our current understanding of the applicability of complex Langevin dynamics at finite density and review the various ways in which the outcome of a complex Langevin process can be assessed, in particular when the exact result is not available.
  Sec.\ \ref{sec:just} constitutes the main part of the paper.  Here we present a variety of numerical results assessing the applicability of complex Langevin dynamics in this model, both in the disordered and the ordered phase. We also demonstrate that the higher-order algorithm eliminates most of the stepsize dependence.
Conclusions are drawn in Sec.~\ref{sec:sum}. The higher-order algorithm is discussed in some more detail in Appendix \ref{sec:appA}, while Appendix \ref{sec:tables} can be used to scrutinize the stepsize dependence and criteria for correctness. A brief account of part of this work has appeared in Ref.\ \cite{pos}.

\section{SU(3) spin model}
\label{sec:model}

We consider the three-dimensional SU(3) spin model at nonzero chemical potential, with the action \cite{KW}
\be
 S = S_B+S_F,
 \ee
 where
 \begin{align}
 S_B  & = -\beta\sum_{x}\sum_{\nu=1}^3 \left(  \Tr U_x\Tr U_{x+\hat\nu}^\dagger + \Tr U_x^\dagger\Tr U_{x+\hat\nu}\right),
 \\
S_F & =  -h \sum_x\left( e^\mu\Tr U_x + e^{-\mu}\Tr U_x^\dagger\right).
\end{align}
The model can be thought of as an effective dimensionally reduced version of QCD, where $\Tr U_x$ represents  the trace of the Polyakov loop; the $U_x$'s are SU(3) matrices living on a three-dimensional lattice (we use periodic boundary conditions). The first term then represents the gluon contribution with effective coupling $\beta$, while the second term represents heavy quarks, with coupling $h$. Chemical potential favours quarks over anti-quarks, resulting in a complex action,  $S_F^*(\mu)=S_F(-\mu^*)$.  The fermion term is a simplified version of the contribution derived in the heavy dense limit \cite{Aarts:2008rr}.
The partition function,
\be
Z = e^{-\Omega f} = \int \prod_x dU_x\, e^{-S},
\ee
is even in $\mu$ due to charge conjugation invariance. 
Here $f$ denotes the free energy density and  $\Omega$ is the three-dimensional volume.

The phase structure of this theory has been  studied in Refs.\ \cite{KW,BGS}, using both complex Langevin dynamics and mean-field theory. 
Recently it has also been investigated using a reformulation of the theory which is sign-problem free \cite{arXiv:1111.0916}.
For small $h$, the theory has a disordered (confined)  phase for smaller $\beta$ values, and an ordered (deconfined) phase for larger $\beta$ values. The two phases are separated by a first-order phase transition. This is the case for vanishing and small chemical potential. With increasing chemical potential, the transition weakens and turns into a crossover at a critical endpoint. For larger $h$, there is a crossover only.

We will also consider two closely related models which have a real action, namely the model with imaginary chemical potential, $\mu=i\mu_\rmI$, and the phase-quenched model, obtained by discarding the imaginary part of the action, such that
\be
S_F^{\rm pq} =  -h\cosh\mu \sum_x\left( \Tr U_x + \Tr U_x^\dagger\right).
\ee

In contrast to QCD, the SU(3) spin model does not have a Silver Blaze problem. The Silver Blaze problem  \cite{Cohen:2003kd} refers to the region in the phase diagram where the chemical potential is nonzero but bulk thermodynamic observables, such as the pressure and the density, are $\mu$-independent.  This $\mu$-independence requires a precise cancelation which can be highly non-trivial in a numerical approach, as can be seen from studies of  the eigenvalues of the Dirac operator \cite{Cohen:2003kd,Osborn:2005ss}. We note here that complex Langevin dynamics has been shown to be able to solve the Silver Blaze problem, in the relativistic Bose gas \cite{Aarts:2008wh,Aarts:2009hn} and in one-dimensional QCD \cite{arXiv:1006.0332}. 
To see that the Silver Blaze region is absent in this model, consider the density,
 \be
\left\bra n\right\ket  = \frac{1}{\Omega}\frac{\partial\ln Z}{\partial\mu} =  \left\bra he^\mu \Tr U_x - h e^{-\mu}\Tr U_x^\dagger \right\ket.
 \ee
  A nonzero density induces a difference between $\bra\Tr U_x\ket$ and $\bra\Tr U_x^\dagger\ket$. On the other hand, 
in the Silver Blaze region, $\bra n\ket=0$ and $\bra\Tr U_x\ket=\bra\Tr U_x^\dagger\ket$. It is clear from the expression above that it is not possible to simultaneously satisfy these conditions when $\mu\neq 0$, hence the Silver Blaze region is absent.\footnote{For completeness, we recall that $\bra\Tr U_x\ket$ and $\bra\Tr U_x^\dagger\ket$ are both real in the full theory, while 
at imaginary $\mu$ the real parts are equal and the imaginary parts are opposite. In the phase-quenched theory,  they are real and identical. Note also that the fermion contribution breaks the Z(3) symmetry of the bosonic sector, hence $\bra\Tr U_x\ket$ and $\bra\Tr U_x^\dagger\ket$ are never strictly zero.}
Similarly, the density in the phase-quenched theory,
 \be
\bra n\ket_{\rm pq}  = h\sinh\mu\left\bra \Tr U_x + \Tr U_x^\dagger\right\ket_{\rm pq},
  \ee
is nonzero as soon as $\mu>0$. 

The severeness of the sign problem is conventionally estimated via the expectation value of the complex phase factor 
$e^{i\varphi} = e^{-S}/|e^{-S}|$ in the phase quenched theory, 
\be
\bra e^{i\varphi}\ket_{\rm pq} = \frac{Z}{Z_{\rm pq}} = e^{-\Omega\left(f-f_{\rm pq}\right)}.
\ee
The full and the phase-quenched theory differ as soon as $\mu$ is nonzero, which can be seen by performing a Taylor series expansion around $\mu=0$. To second order in $\mu$, the free energy densities read
\begin{align}
\label{eq:f}
f(\mu) & =  f(0) -   \left(c_1+ c_2h\right) h\mu^2  +{\cal O}(\mu^4), \\
f_{\rm pq}(\mu) & =  f(0)- c_1h\mu^2 + {\cal O}(\mu^4),
\label{eq:fpq}
\end{align}
with
\begin{align}
\label{eq:c1}
c_1 & = \frac{1}{\Omega}  \sum_x\left\bra\Tr U_x\right\ket_{\mu=0}, \\
\label{eq:c2}
c_2 & =   \frac{1}{2\Omega} \sum_{xy} \left\bra \Trr\left(U_x-U_x^\dagger\right) \Trr\left(U_y-U_y^\dagger\right) \right\ket_{\mu=0}.
\end{align}
Since $c_2$ is negative [$\Trr\left(U_x-U_x^\dagger\right)$ is imaginary],  $f - f_{\rm pq}\ge 0$, as it should be. Similarly, $\bra n\ket  \le \bra n\ket_{\rm pq}$.

\section{Discretized complex Langevin dynamics}
\label{sec:dcl}

In order to solve the complex Langevin evolution numerically, we follow Ref.\ \cite{KW} and diagonalize the SU(3) matrices in terms of the angles $\phi_{1,2}$, such that
 \begin{align}
 \Tr U_x & = e^{i\tht_{1x}} + e^{i\tht_{2x}} + e^{-i(\tht_{1x}+\tht_{2x})}, \\
 \Tr U_x^\dagger & = e^{-i\tht_{1x}} + e^{-i\tht_{2x}} + e^{i(\tht_{1x}+\tht_{2x})}.
 \end{align}
 We then have to include the reduced Haar measure and consider the partition function
 \be
 Z = \prod_x \int_{-\pi}^\pi d\phi_{1x} \, d\phi_{2x}\, e^{-S_{\rm eff}},
 \ee
 where
 \be
 S_{\rm eff} = S_B+S_F+S_H,
 \ee
 with
 \be
S_H  = -\sum_x\ln\left[ 
\sin^2\left(\frac{\tht_{1x}-\tht_{2x}}{2}\right)
\sin^2\left(\frac{2\tht_{1x}+\tht_{2x}}{2}\right)
\sin^2\left(\frac{\tht_{1x}+2\tht_{2x}}{2}\right)
\right].
 \ee
We note that it is also possible to implement complex Langevin dynamics directly for the SU(3) matrices, 
see e.g.\ Refs.\ \cite{BGS,Aarts:2008rr}.
 
Langevin dynamics provides a stochastic update for the angles $\phi_{ax}$ ($a=1,2$), according to
 \be
 \label{eq:RL}
\frac{\partial}{\partial \vartheta} \phi_{ax} = K_{ax} +\eta_{ax},
\quad\quad\quad
K_{ax} = -\frac{\partial S_{\rm eff}}{\partial\tht_{ax}},
\ee
where $\vartheta$ denotes the Langevin time, $K_{ax}$ is the drift term, and the noise satisfies
 \be
 \bra \eta_{ax}\ket=0, 
 \quad\quad\quad
 \bra \eta_{ax}\eta_{a'x'}\ket=2\delta_{aa'}\delta_{xx'}.
 \ee

When the action and hence the drift terms are complex, the angles do not remain real under the Langevin evolution. We therefore write
 $\tht_{ax}=\thtR_{ax}+i\thtI_{ax}$, and consider the  following complex Langevin equations, using real noise,
   \begin{align}
&\frac{\partial}{\partial \vartheta} \thtR_{ax} = K^\rmR_{ax} +\eta_{ax},
& \quad\quad
K^\rmR_{ax} = -\re\frac{\partial S_{\rm eff}}{\partial\tht_{ax}} \Big|_{\tht_{ax}\to \thtR_{ax}+i\thtI_{ax}},
\\
&\frac{\partial}{\partial \vartheta} \thtI_{ax} = K^\rmI_{ax},
& \quad\quad
K^\rmI_{ax} = -\im\frac{\partial S_{\rm eff}}{\partial\tht_{ax}} \Big|_{\tht_{ax}\to \thtR_{ax}+i\thtI_{ax}}.
\end{align}
 After complexification,  we write $U_x^{-1}$ instead of $U_x^\dagger$ in the remainder.

To solve these equations numerically, Langevin time is discretized as $\vartheta=\eps n$, where $\eps$ is the Langevin time step. The standard algorithm discretizing Eq.\ (\ref{eq:RL}) reads\footnote{Complexification is obvious and we do not give the discretized equations explicitly.}
\be
\label{eq:sslo}
\phi_{ax}(n+1) = \phi_{ax}(n) +\eps K[\phi_{ax}(n)] +\sqrt\eps\,\eta_{ax}(n),
\ee
where 
\be
\bra\eta_{ax}(n)\ket=0, \quad\quad\quad
\bra\eta_{ax}(n)\eta_{a'x'}(n')\ket=2\delta_{aa'}\delta_{xx'}\delta_{nn'}.
\ee
The contribution to the drift term from the Haar measure requires careful integration. For this we use the adaptive stepsize algorithm of Ref.\ \cite{arXiv:0912.0617}.

It is well-known that Langevin dynamics has finite stepsize corrections, which are linear in $\eps$ in the lowest-order discretization given above \cite{Damgaard:1987rr}.  It is therefore necessary to extrapolate to zero stepsize. In our previous work 
\cite{Aarts:2008rr,Aarts:2008wh,Aarts:2009hn,arXiv:0912.0617,arXiv:0912.3360,arXiv:1005.3468,arXiv:1006.0332,arXiv:1101.3270}, 
we have only considered the lowest-order algorithm. However, motivated by the results to be presented below,  we implemented a  higher-order algorithm to improve the stepsize dependence. A standard Runge-Kutta scheme, where the drift terms are improved but the noise is kept as above, will not remove the leading stepsize correction \cite{kloeden}. Instead, it is necessary to modify the noise terms as well. We use the algorithm proposed in Ref.\ \cite{CCC} for real Langevin dynamics, which is explicit and easy to implement.\footnote{For other approaches, see e.g.\ Refs.\ \cite{Drummond:1982sk,Catterall:1990qn}.}
 It takes the following form
\begin{align}
\nn
\psi_{ax}(n)  		=\; & \phi_{ax}(n) +\half\eps K[\phi_{ax}(n)], \\
\nn
\tilde \psi_{ax}(n) 	=\; & \phi_{ax}(n) +\half\eps K[\phi_{ax}(n)] + \frac{3}{2}\sqrt{\eps}\,\tilde\alpha_{ax}(n), \\
\phi_{ax}(n+1) 		=\; & \phi_{ax}(n) + \frac{1}{3} \eps \left( K[\psi_{ax}(n)] + 2 K[\tilde \psi_{ax}(n)]\right) +\sqrt\eps\,\alpha_{ax}(n).
\label{eq:CCC}
\end{align}
Here $\tilde\alpha_{ax}(n)$ is a random variable taken according to
\be
\tilde\alpha_{ax}(n) = \half\alpha_{ax}(n)+\frac{\sqrt 3}{6}\xi_{ax}(n),
\ee
while $\alpha_{ax}(n)$ and $\xi_{ax}(n)$ are independent Gaussian random variables with variance 2 and vanishing mean, i.e.,
\begin{align}
\nn
& \bra\alpha_{ax}(n)\alpha_{a'x'}(n')\ket = \bra\xi_{ax}(n)\xi_{a'x'}(n')\ket = 2\delta_{aa'}\delta_{xx'}\delta_{nn'}, 
\\
& \bra\alpha_{ax}(n)\xi_{a'x'}(n')\ket = \bra\alpha_{ax}(n)\ket =  \bra\xi_{ax}(n)\ket = 0. 
\end{align}
In Ref.\ \cite{CCC} it was shown analytically, for the case of a real drift term, that with this update the remaining correction is  ${\cal O}(\eps^2)$ 
for a system with one degree of freedom and  ${\cal O}(\eps^{3/2})$ for a coupled system. 
In Appendix \ref{sec:appA} we discuss this algorithm in some more detail.

\section{Justification and criteria for correctness}
\label{sec:cfc}

In the case of a real action, it can be shown that stochastic quantization and Langevin dynamics is equivalent to standard path integral quantization  \cite{Damgaard:1987rr}.
As is well-known, such a general statement is lacking in the case of a complex action \cite{Damgaard:1987rr}. Indeed, it can occur that under complex Langevin evolution expectation values converge to a wrong result 
\cite{Klauder:1985b,Ambjorn:1985iw,Ambjorn:1986fz,SI-91-8,Berges:2006xc,Berges:2007nr,arXiv:1005.3468}.
 It is therefore important to be able to judge the outcome of a complex Langevin process  using assessments which are general and can be used in a variety of theories, especially when there are no known results to compare with. 

The first assessment employs analyticity at small $\mu^2$: observables, which are even under charge conjugation, should be analytic as a function of $\mu^2$ (in a finite volume) \cite{Lombardo:1999cz,deForcrand:2002ci}. Results at positive $\mu^2$ can be 
compared with those where $\mu^2<0$, i.e.\ at imaginary potential, obtained using real Langevin dynamics or any other standard approach. This test is limited to small chemical potentials.

A more formal justification of the complexified dynamics can be found in Refs.~\cite{arXiv:0912.3360,arXiv:1101.3270}. Here we summarize that discussion briefly in order to arrive at the criteria for correctness developed in Ref.\ 
\cite{arXiv:1101.3270}. We consider expectation values with respect to the real and positive probability distribution $P[\phi^\rmR, \phi^\rmI;\vartheta]$, sampled by the stochastic process,
\be
\label{eq:42}
\bra O\ket_{P(\vartheta)}  = \frac{\int D\phi^\rmR D\phi^\rmI\, P[\phi^\rmR, \phi^\rmI;\vartheta]O[\phi^\rmR+i\phi^\rmI]}
                                  {\int D\phi^\rmR \phi^\rmI\, P[\phi^\rmR, \phi^\rmI;\vartheta]}.
\ee
Here $\vartheta$ is the Langevin time. With the help of the Langevin equations for $\phi^\rmR$ and $\phi^\rmI$, one finds the Fokker-Planck equation for $P[\phi^\rmR, \phi^\rmI;\vartheta]$,
\be 
\label{eq:44}
\frac{\partial P[\phi^\rmR, \phi^\rmI;\vartheta]}{\partial\vartheta} = L^TP[\phi^\rmR, \phi^\rmI;\vartheta],
\quad\quad
L^T = \frac{\partial}{\partial\phi^\rmR}\left[\frac{\partial}{\partial\phi^\rmR} -K^\rmR\right] -  \frac{\partial}{\partial\phi^\rmI} K^\rmI.
\ee
We also consider expectation values with respect to a complex weight $\rho[\phi;\vartheta]$,
\be
\label{eq:41}
\bra O\ket_{\rho(\vartheta)} = \frac{\int D\phi\, \rho[\phi;\vartheta]O[\phi]}{\int D\phi\, \rho[\phi;\vartheta]},
\ee
where $\phi$ is real  and $\rho[\phi;\vartheta]$ satisfies 
\be 
\label{eq:43}
\frac{\partial\rho[\phi;\vartheta]}{\partial\vartheta} = L_0^T\rho[\phi;\vartheta], 
\quad\quad\quad
L_0^T = \frac{\partial}{\partial\phi}\left[\frac{\partial}{\partial\phi} + \frac{\partial S}{\partial\phi}\right].
\ee 
Note that the Fokker-Planck operators $L^T$,  acting on the real density $P[\phi^\rmR, \phi^\rmI;\vartheta]$,
and $L_0^T$, acting on the complex density $\rho[\phi;\vartheta]$, should be distinguished. Furthermore, 
 Eq.\ (\ref{eq:43}) has a stationary solution, $\rho[\phi]\sim \exp(-S)$, whereas for  Eq.\ (\ref{eq:44}) no generic stationary solution is known.

Employing that the only permissible observables are holomorphic and making use of partial integration, one can show that expectation values with respect to the two densities are equal,
\be
\label{eq:45}
\bra O\ket_{P(\vartheta)}   =  \bra O\ket_{\rho(\vartheta)}.
\ee
If subsequently one can show that $\rho[\phi;\vartheta]$ reaches the unique stationary solution $\sim \exp(-S)$  in the limit that $\vartheta\to\infty$,
 the use of complex Langevin dynamics is justified \cite{arXiv:0912.3360}.

The equivalence in Eq.\ (\ref{eq:45}) relies on the ability to do partial integration without receiving contributions from boundary terms, i.e.\ the distributions should be well localized and decay strongly. It was shown in Ref.\ \cite{arXiv:1101.3270} that this condition can be expressed as a  set of criteria on holomorphic observables $O$, which take the form 
\be
\label{eq:LO}
\bra \tilde L O\ket = 0.
\ee
Here $\tilde L$ denotes the Langevin operator
\be
\tilde L =  \left( \frac{\partial}{\partial \phi}  + K\right) \frac{\partial}{\partial \phi},
\quad\quad\quad
K=-\frac{\partial S}{\partial\phi},
\ee
which depends on holomorphic degrees of freedom $\phi$. Although it differs from $L$ and $L_0$, the action of $\tilde L$ on holomorphic observables agrees with that of $L$. The expectation value in Eq.\ (\ref{eq:LO}) is taken with respect to the weight $P$ in the limit that the Langevin process has equilibrated ($\vartheta\to\infty$).
In principle, the criteria (\ref{eq:LO}) should be satisfied for a large enough set of holomorphic observables  \cite{arXiv:1101.3270}.

Adapting this to the model at hand, $\tilde L$ reads
\be
\tilde L = \sum_{x,a} \left( \frac{\partial}{\partial \tht_{ax}}  + K_{ax}\right) \frac{\partial}{\partial \tht_{ax}}.
\ee
We will consider only local observables and denote these as $O[\tht_{1x},\tht_{2x}] = O_x$. We can then write
\be
\tilde L O_x  = \sum_a \left( O^{a''}_x + K_{ax} O_x^{a'} \right),
\ee
 where
 \be
  O_x^{a'} = \frac{\partial O_x}{\partial \tht_{ax}}, 
  \;\;\;\;\;\;\;\;
  O_x^{a''} = \frac{\partial^2 O_x}{\partial \tht_{ax}^2}.
  \ee
  In terms of the real and imaginary parts, this yields explicitly
\begin{align}
 \re \tilde L O_x  = & \sum_a \left( \re O^{a''}_x + K_{ax}^\rmR \re O_x^{a'} -   K_{ax}^\rmI \im O_x^{a'} \right), \\
 \im \tilde L O_x  = & \sum_a \left( \im O^{a''}_x + K_{ax}^\rmR \im O_x^{a'} +   K_{ax}^\rmI \re O_x^{a'} \right),
 \end{align}
 and the criteria read
 \be
 \bra  \re \tilde L O_x\ket = 0,
  \quad\quad\quad
   \bra  \im \tilde L O_x\ket = 0.
   \ee

\section{Results and justification} 
\label{sec:just}

In this section we present a number of results obtained with complex Langevin dynamics. As mentioned earlier, our goal is not deliver a detailed study of critical properties and the phase structure; rather the aim of this study is to assess the reliability of the complex  Langevin algorithm, using the criteria discussed in the previous section.

\begin{figure}[h]
 \begin{center}
  \includegraphics[width=0.7\textwidth]{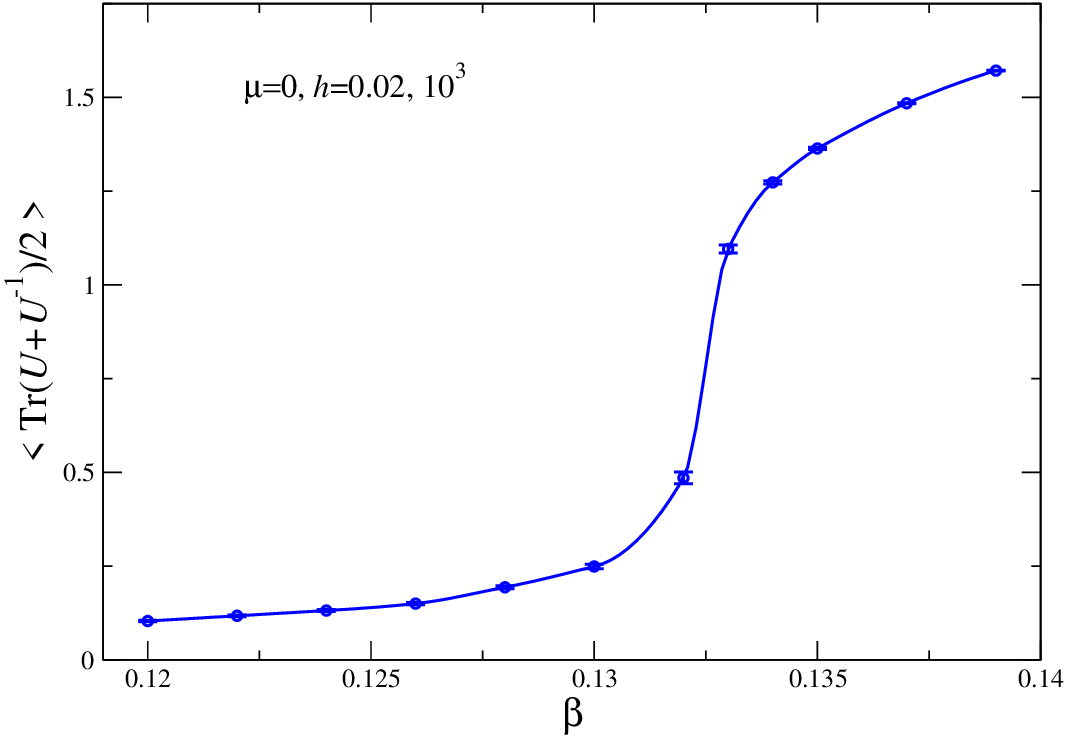}  
 \end{center}
 \caption{$\bra \Tr \left(U_x+U_x^{-1}\right)/2\ket$ as a function of $\beta$ at $\mu=0$ and $h=0.02$ on a $10^3$ lattice, using real Langevin dynamics.}
\label{fig:mu0}
\end{figure}

The results we show here are obtained using a relatively small value for the fermion coupling, $h=0.02$, so that there is a clear transition between the ordered and the disordered phase. We consider $\beta$ values between 0.12 and 0.139; the critical $\beta$ value for the fermion coupling we use is around 0.1324 at $\mu=0$. This is demonstrated in Fig.\ \ref{fig:mu0}, where we show $\bra\Trr\left(U_x+U_x^{-1}\right)/2\ket = \bra\Tr U_x\ket$ as a function of $\beta$ at $\mu=0$ on a $10^3$ lattice.

\begin{figure}[!t]
 \begin{center}
 \includegraphics[width=0.7\textwidth]{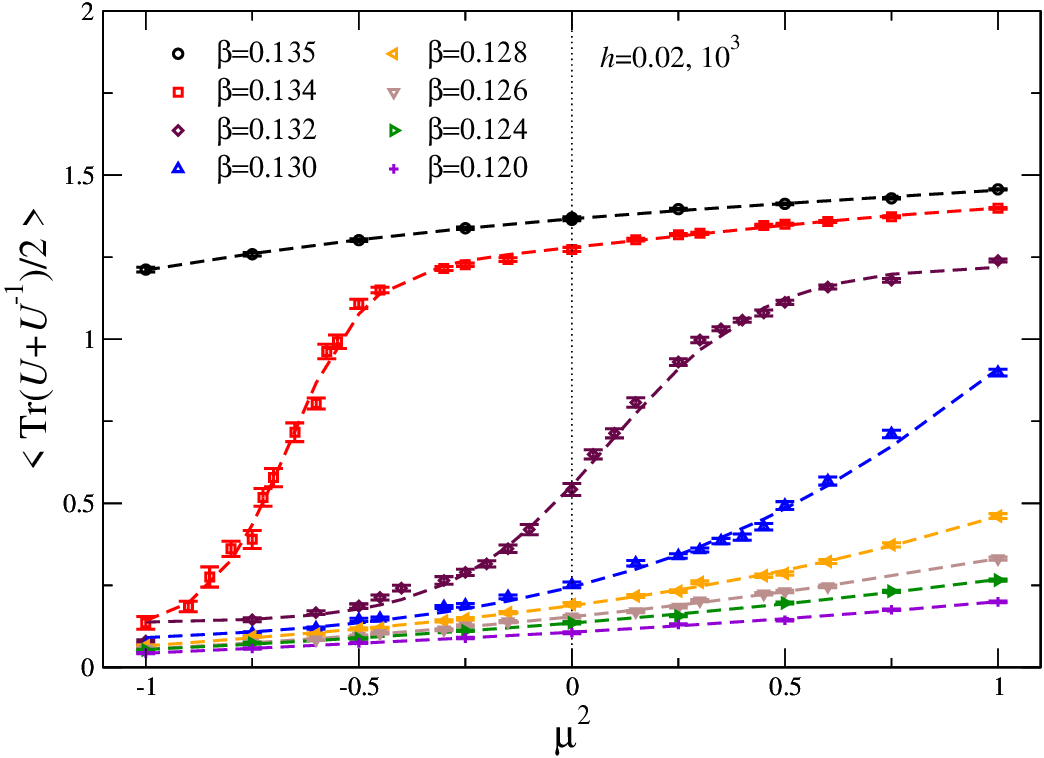} 
  \end{center}
 \caption{Analyticity in $\mu^2$: $\left\bra\Trr\left (U_x+U_x^{-1}\right)/2\right\ket$ as a function of $\mu^2$ for various $\beta$ values with $h=0.02$ on a $10^3$ lattice. Data at imaginary $\mu$ (with $\mu^2\le 0$) has been obtained with real Langevin dynamics, data at real $\mu$ (with $\mu^2> 0$)  with complex Langevin dynamics.}
\label{fig:ana}
\end{figure}

As a first test, we probe the transition by varying $\mu$ instead of $\beta$. As mentioned above, 
observables which are invariant under charge conjugation should, in a finite volume, be analytic in $\mu^2$
\cite{Lombardo:1999cz,deForcrand:2002ci}. This yields the possibility to 
compare results at positive $\mu^2$ with those at negative $\mu^2$, corresponding to imaginary potential. Since in this case the action is real,  real Langevin dynamics can be used, which is theoretically well founded. 
 In Fig.\ \ref{fig:ana} we show  $\left\bra\Trr\left(U_x+U_x^{-1}\right)/2\right\ket$ as a function of $\mu^2$ for eight different $\beta$ values. 
  We observe smooth behaviour as $\mu^2$ is increased. This is an indication that complex Langevin dynamics works well. We note that this is true in both phases as well as in the transition region. 
This is in contrast to the case of the XY model recently studied using complex Langevin dynamics, where correct results were obtained in only part of the phase diagram \cite{arXiv:1005.3468}.

\begin{figure}[!t]
 \begin{center}
  \includegraphics[width=0.49\textwidth]{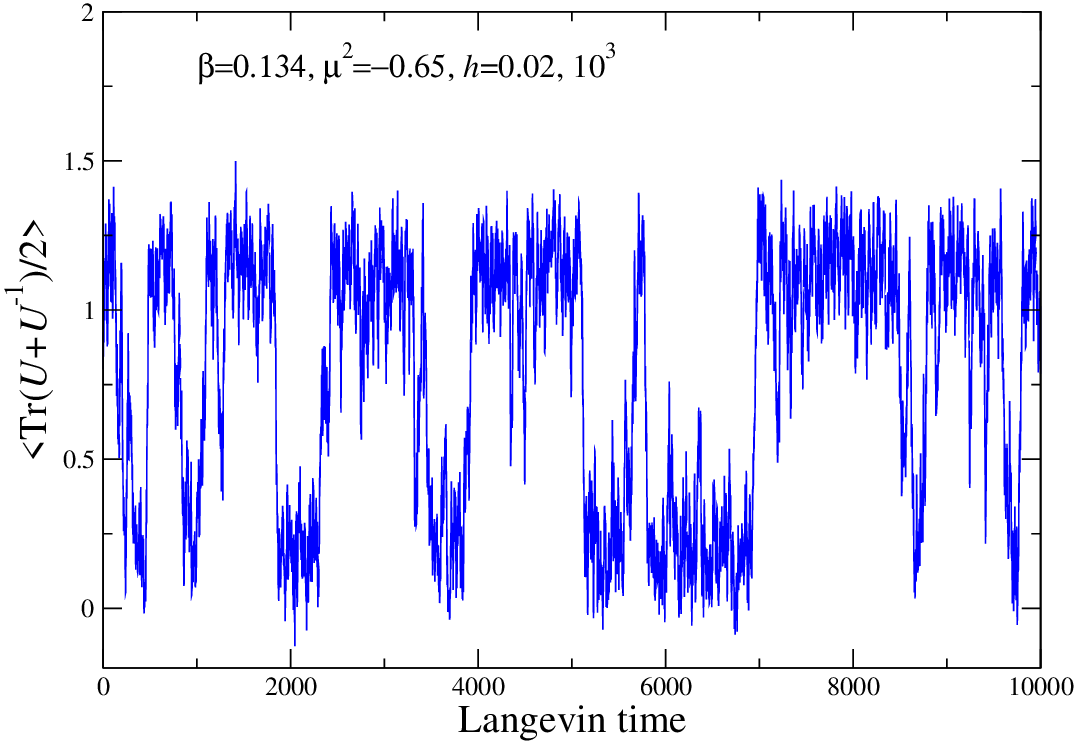} 
  \includegraphics[width=0.49\textwidth]{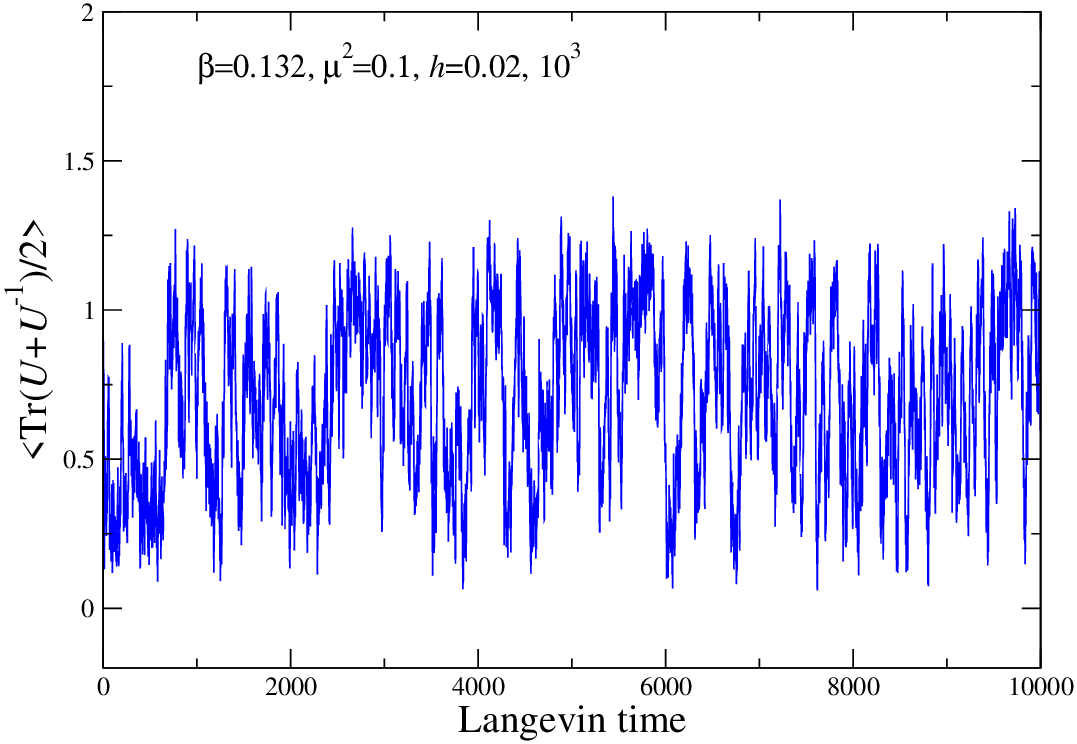} 
 \end{center}
 \caption{Langevin time evolution of $\bra\Trr\left(U_x+U_x^{-1}\right)/2\ket$ in the transition region, at imaginary chemical, $\mu^2=-0.65$ and $\beta=0.134$ (left) and real chemical potential, $\mu^2=0.1$ and $\beta=0.132$ (right). The other parameters are as above.
 }
\label{fig:hist}
% \end{figure}
%\begin{figure}[t]
 \begin{center}
 \includegraphics[width=0.7\textwidth]{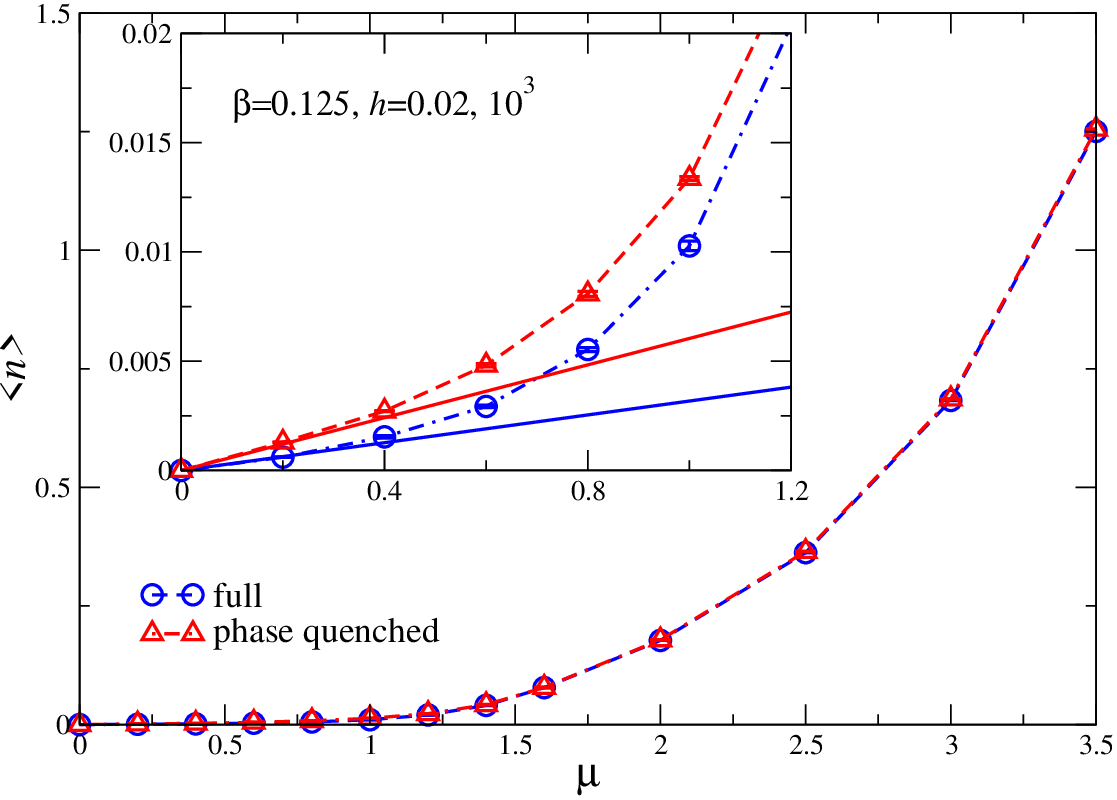} 
\end{center}
 \caption{Density $\bra n\ket$ in the full and the phase-quenched theory as a function of $\mu$ at $\beta=0.125$ and $h=0.02$ on a $10^3$ lattice. The inset shows a close-up of the small $\mu$ region. The lines are the predicted linear dependence for small $\mu$, evaluated at $\mu=0$.
  } 
\label{fig:dens}
\end{figure}

The strength of the transition weakens as $\mu^2$ increases and, vice versa, increases as $\mu^2$ decreases
\cite{KW,BGS,arXiv:1111.0916}. This can be seen in Fig.\ \ref{fig:hist}, where we show the Langevin time evolution of $\left\bra\Trr\left(U_x+U_x^{-1}\right)/2\right\ket$ at $\mu^2=-0.65$  and $\beta=0.134$ (left) and $\mu^2=0.1$  and $\beta=0.132$ (right).
We observe clear first order behaviour at $\mu^2=-0.65$, while at $\mu^2=0.1$ the transition is much weaker.

Next we consider the density as a function of $\mu$ in the full and the phase-quenched theory. In Fig.\ \ref{fig:dens} we show the density for chemical potentials up to $\mu=3.5$, at $\beta=0.125$. For this $\beta$ value the model is in the disordered phase at smaller $\mu$ values and in the ordered phase at larger $\mu$ values.
The densities in the full and phase-quenched theories are similar, but not equal.
We recall that there is no Silver Blaze region in this model. This can be seen in the inset, which shows a close-up: 
the density in the full theory is below the one in the phase-quenched theory, but it is nonzero (we have verified that there are no visible finite-size effects remaining).
The lines indicate the expected linear dependence of the densities on $\mu$, using the lowest-order Taylor series expansion, 
see Eqs.~(\ref{eq:f}, \ref{eq:fpq}),
\be
\bra n\ket =  2\left(c_1+c_2h\right)h\mu +{\cal O}(\mu^3),
 \ee
 where the coefficients $c_{1,2}$ have been defined in Eqs.\ (\ref{eq:c1}, \ref{eq:c2}).
 In the phase-quenched theory the term with $c_2$ is absent. We have computed the coefficients and find
\be
\label{eq:ll}
c_1 = 0.1446(21),
\quad\quad\quad
c_2 = -3.534(72).
\ee
Using these coefficients yields the straight lines in the inset of Fig.\ \ref{fig:dens}, justifying the results of complex Langevin dynamics.

\begin{figure}[t]
 \begin{center}
    \includegraphics[width=0.7\textwidth]{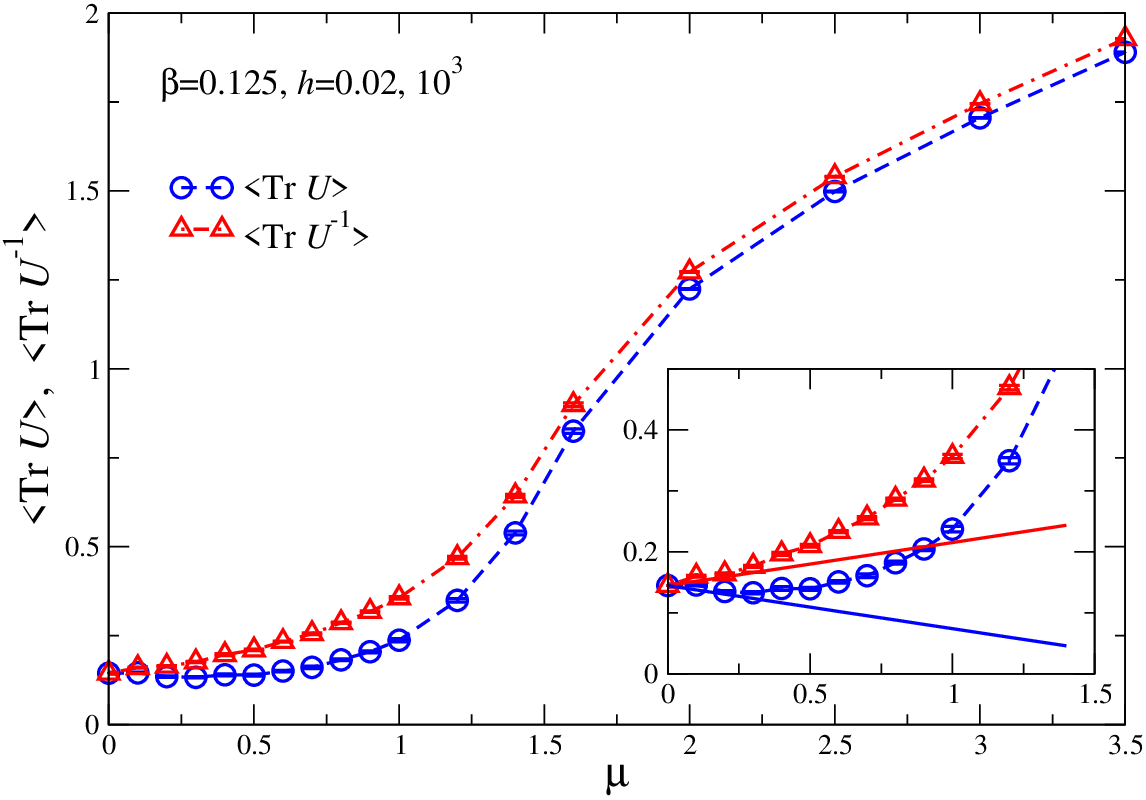} 
 \end{center}
 \caption{$\bra\Tr U_x\ket$ and  $\bra\Tr U_x^{-1}\ket$ as a function of $\mu$ in the full theory.
 The parameters are as in Fig.\ \ref{fig:dens}. The inset shows a close-up of the small $\mu$ region. The lines are the predicted linear dependence for small $\mu$, evaluated at $\mu=0$.
  }
\label{fig:trutruinv}
 \end{figure}

In Fig.\ \ref{fig:trutruinv} we show $\bra\Tr U\ket$ and  $\bra\Tr U^{-1}\ket$ as a function of $\mu$ in the full theory, using the same parameters as in Fig.\  \ref{fig:dens}. Recall that $\bra\Tr U\ket$ and  $\bra\Tr U^{-1}\ket$ are both real and that one expects
$\bra\Tr U\ket < \bra\Tr U^{-1}\ket$, due to the nonzero density. At small $\mu$, the linear dependence on $\mu$ can again be expressed in terms of the coefficients $c_{1,2}$ and we find
\begin{align}
\bra\Tr U \ket & = c_1 + c_2h\mu + {\cal O}(\mu^2), \\
\bra\Tr U^{-1} \ket & = c_1 - c_2h\mu + {\cal O}(\mu^2).
\end{align}
This yields the straight lines in the inset of Fig.\ \ref{fig:trutruinv}, justifying again the results of complex Langevin dynamics. In the phase-quenched theory, $\bra \Tr U\ket_{\rm pq}$ and $\bra \Tr U^{-1}\ket_{\rm pq}$  are equal and slightly below 
 $\bra \Trr\left(U+U^{-1}\right)/2\ket$ in the full theory (not shown).

\begin{figure}[t]
 \begin{center}
  \includegraphics[width=0.7\textwidth]{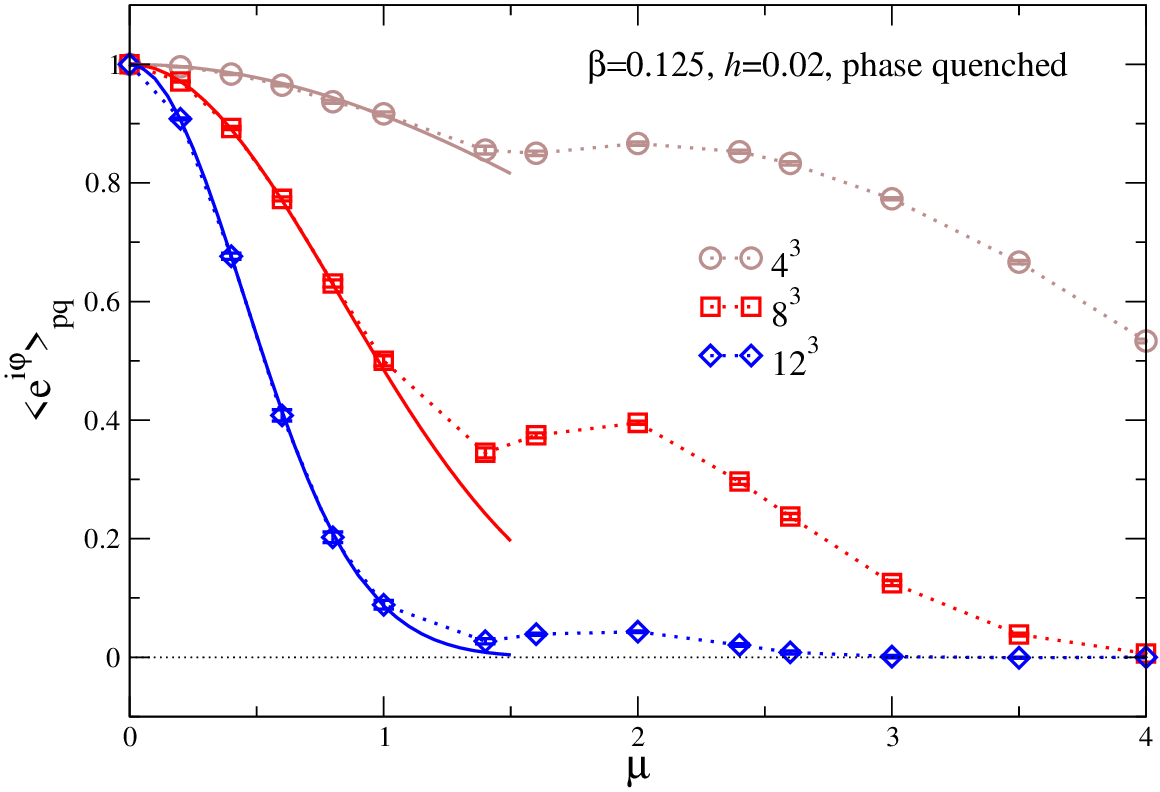} 
 \end{center}
 \caption{
Average phase factor in the phase-quenched theory $\bra e^{i\varphi}\ket_{\rm pq}$ as a function of $\mu$, for various volumes at $\beta=0.125$ and $h=0.02$. The lines indicates the expected behaviour  using the leading $\mu^2$ term at small $\mu$.
} 
\label{fig:avsign}
 \end{figure}

The average phase factor, indicating the severeness of the sign problem, is shown in Fig.\ \ref{fig:avsign} for a typical choice of parameters. 
The lines indicate the behaviour expected at small chemical potential,
\be
 \bra e^{i\varphi}\ket_{\rm pq}  = e^{-\Omega\Delta f}, 
 \quad\quad\quad
\Delta f = f-f_{\rm pq} = -c_2h^2\mu^2 +{\cal O}(\mu^4).
\ee
As in preceding studies \cite{Aarts:2008rr,Aarts:2008wh,Aarts:2009hn,arXiv:1005.3468,arXiv:1006.0332}, 
we have not observed a correlation between
the severeness of the sign problem and the efficiency of the complex Langevin algorithm.
We also note that the average phase factor behaves in a non-monotonic manner as a function of $\mu$ in the transition region.

In order  to assess complex Langevin dynamics in detail for larger $\mu$ values, we now focus on two points in the phase diagram: $\beta=0.125$, $\mu=1$ in the disordered phase and  $\beta=0.125$, $\mu=3$ in the ordered phase. To control  the statistical error we have carried out simulations using a total Langevin trajectory of length 10,000 (after discarding the thermalization stage) in the disordered phase; in the ordered 
phase fluctuations are smaller and a Langevin trajectory of 5,000 is sufficient. Errors are determined with a jackknife analysis.
We have used a number of stepsizes, from $\eps=0.001$ down to $\eps=0.00005$, employing both the standard lowest-order algorithm and the improved higher-order algorithm. The results are collected in Tables \ref{tab:hpmu1} and \ref{tab:hpmu3} in Appendix \ref{sec:tables}.

In Fig.\ \ref{fig:ss1} we show $\bra\Tr U\ket$ and $\bra\Tr U^{-1}\ket$ as a function of the Langevin stepsize for $\mu=1$ (left) and 3 (right).
Statistical fluctuations in the disordered phase are larger,  even though the Langevin trajectory is twice as long.
 For the lowest-order algorithm stepsize dependence is clearly visible, as expected. The dotted lines indicate a linear fit using the data at the four smallest stepsizes. In the case of the higher-order algorithm, there appears  to be no stepsize dependence visible; the dashed lines indicate the average of the five data points in each case.\footnote{Theoretically, corrections of ${\cal O}(\eps^{3/2})$ are expected \cite{CCC}.} 
 Importantly, we note that the results from both algorithms are consistent in the limit $\eps\to 0$, see also Appendix \ref{sec:tables}.

\begin{figure}[!t]
 \begin{center}
  \includegraphics[width=0.49\textwidth]{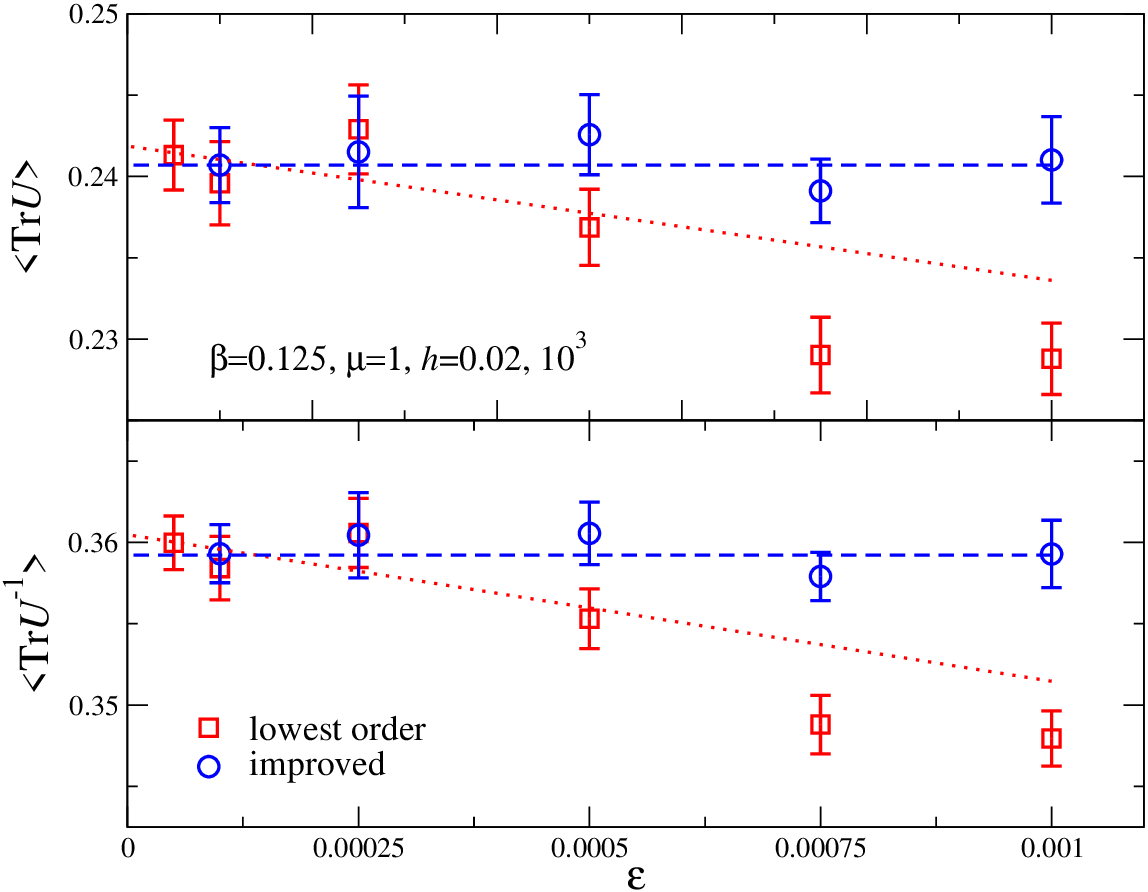} 
  \includegraphics[width=0.49\textwidth]{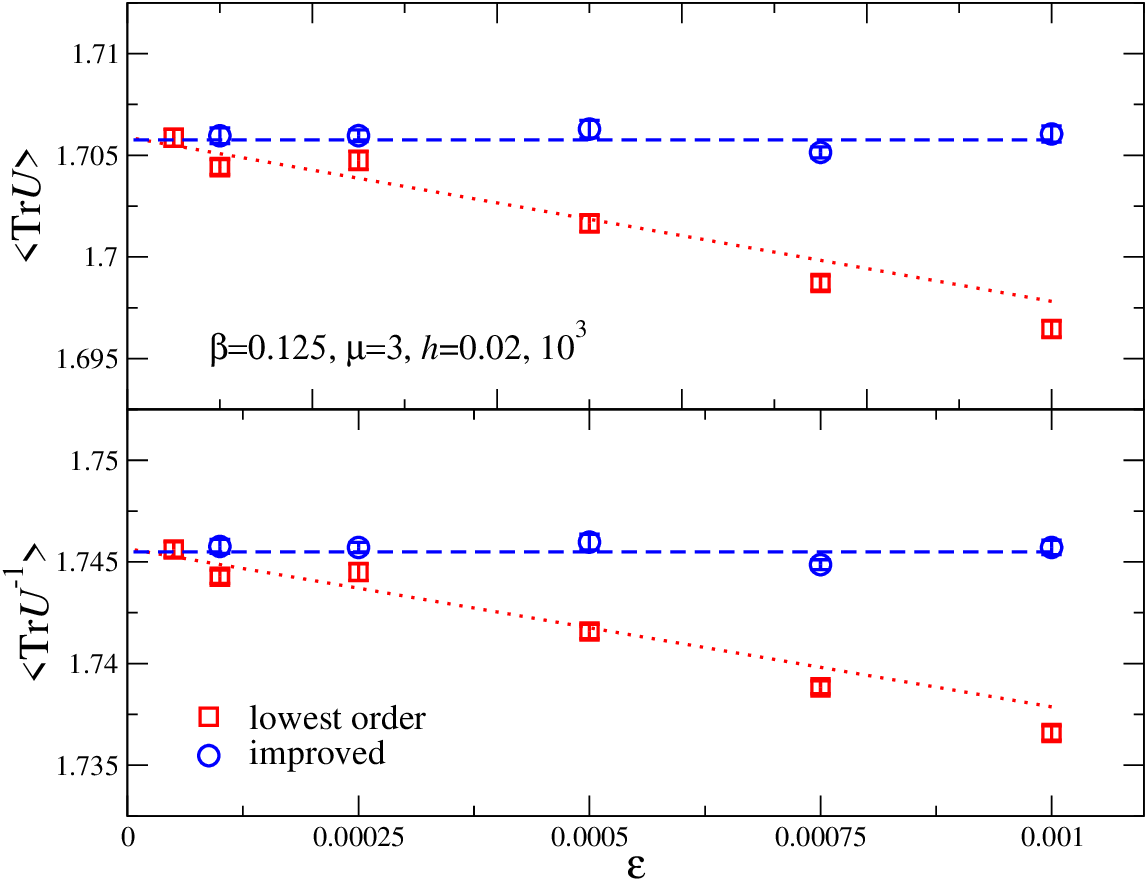} 
 \end{center}
 \caption{Stepsize dependence of $\bra \Tr U_x\ket$ (top panes) and $\bra\Tr U_x^{-1}\ket$ (bottom panes) at $\mu=1$ (left) and 3 (right) on a $10^3$ lattice for $\beta=0.125$ and $h=0.02$,  using both the standard lowest-order and the improved algorithm.     }
\label{fig:ss1}
% \end{figure}
%\begin{figure}[h]
 \begin{center}
  \includegraphics[width=0.49\textwidth]{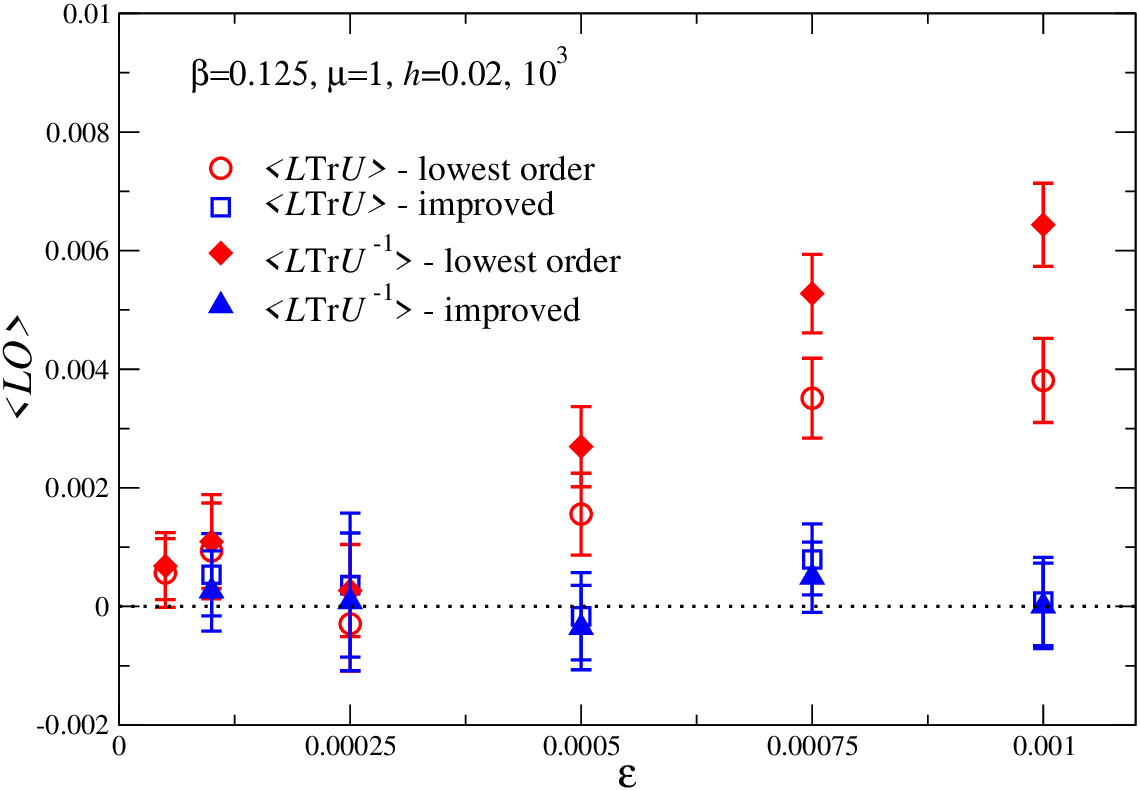} 
  \includegraphics[width=0.49\textwidth]{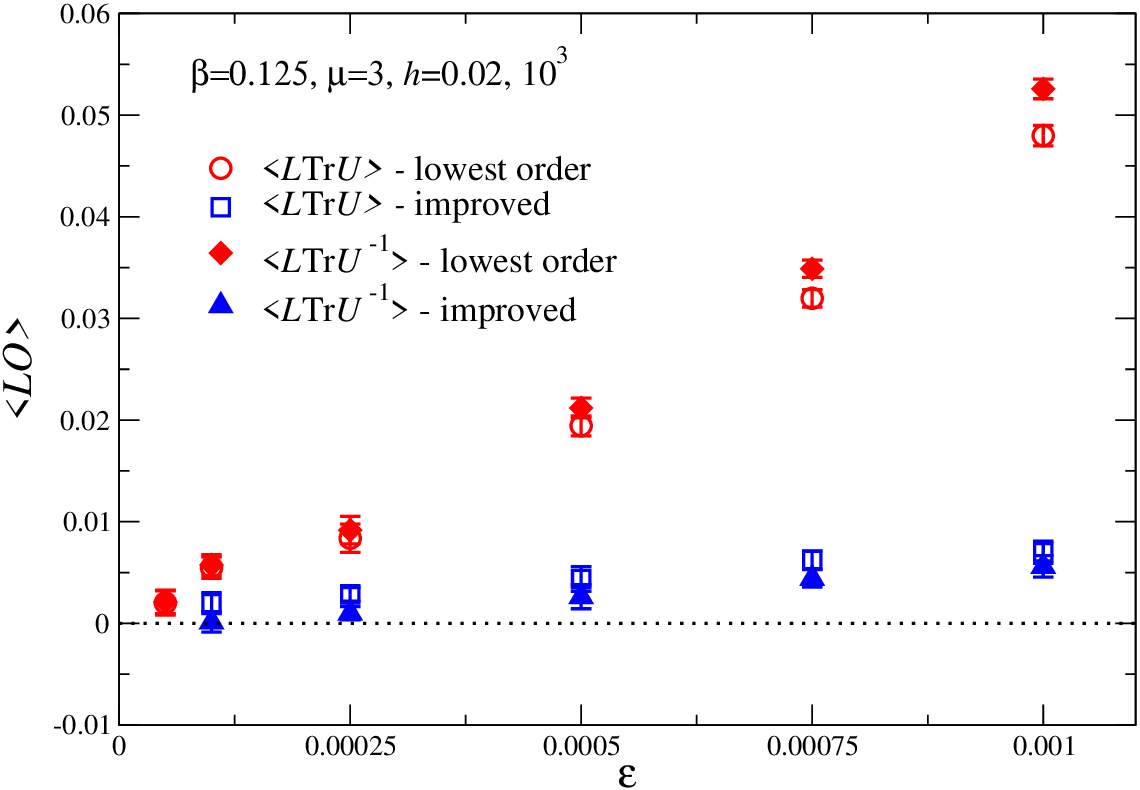} 
 \end{center}
 \caption{Stepsize dependence of the real part of  $\bra L\Tr U_x\ket$ and $\bra L\Tr U_x^{-1}\ket$ at $\mu=1$ (left) and 3 (right), using both the standard and the improved algorithm. Other parameters as in Fig.\ \ref{fig:ss1}.
    }
\label{fig:ss2}
 \end{figure}

In order to justify these results, we have computed $\bra\tilde L O\ket$ (which should be equal to zero),
where $O=\Tr U, \Tr U^{-1}$ and $n$. Since these observables are holomorphic, we drop the tilde on the $L$ from now on.
Note that $\bra L n\ket$ is not independent, since $n$ is a linear combination of $\Tr U$ and $\Tr U^{-1}$.
 The imaginary parts of $\bra L O\ket$ are  consistent with zero. The stepsize dependence of the real parts is shown in Fig.~\ref{fig:ss2} for $\mu=1$ (left) and 3 (right). Note the different vertical scale: the stepsize dependence is stronger  in the ordered phase.\footnote{Larger stepsize corrections in the ordered phase at larger values of $\mu$ are also seen in the Bose gas, where the stepsize is effectively enhanced as $e^\mu\eps$ \cite{Aarts:2009hn}.}
For the lowest-order algorithm there are again clear finite-stepsize corrections, which vanish in the limit that $\eps\to 0$. In the case of the higher-order algorithm,  finite-stepsize corrections are much smaller or even absent.
 We find that $\bra LO\ket$ goes to zero in the limit that the stepsize is taken to zero. This observation is a necessary requirement for the applicability of  complex Langevin dynamics. Interestingly, larger finite-stepsize corrections correspond to larger deviations of $\bra LO\ket$ from zero. It turns out that this is also seen when using real Langevin dynamics, e.g.\ in the  phase-quenched theory. We conclude therefore that computations of $\bra LO\ket$ yield a sensitive test to  quantify finite-stepsize errors.

\begin{figure}[!t]
 \begin{center}
  \includegraphics[width=0.49\textwidth]{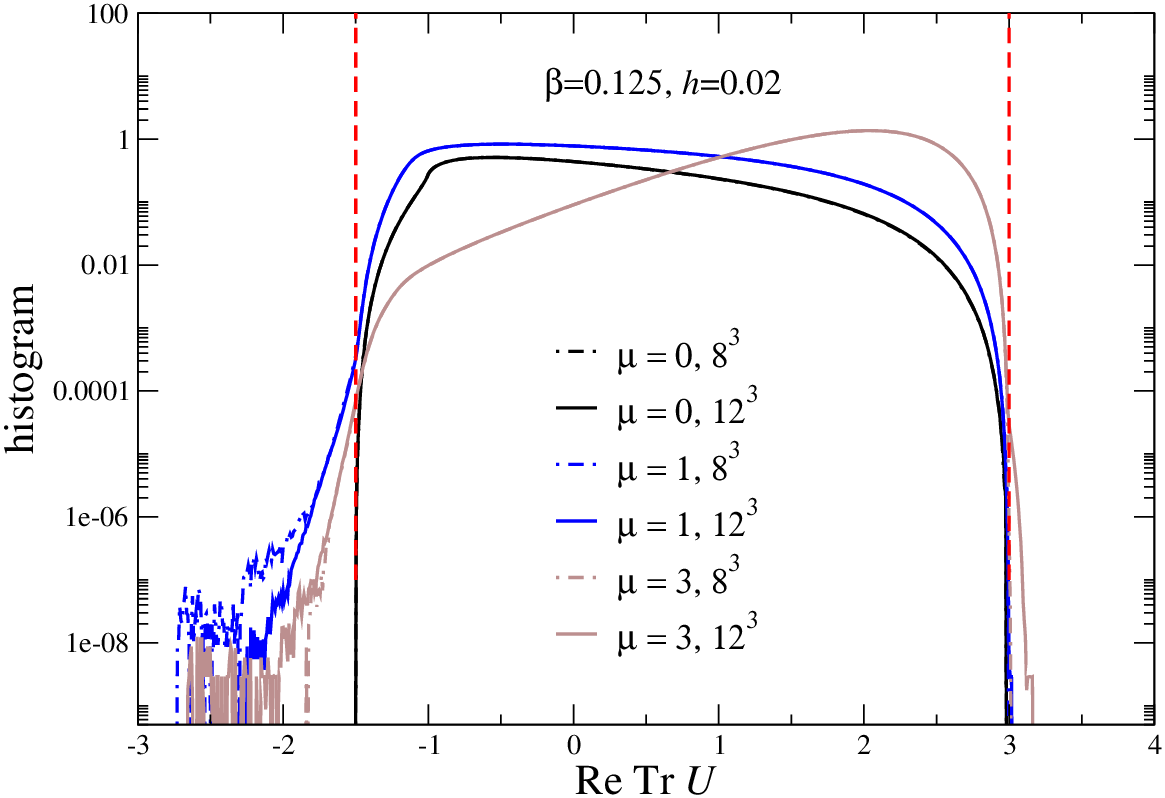} 
  \includegraphics[width=0.49\textwidth]{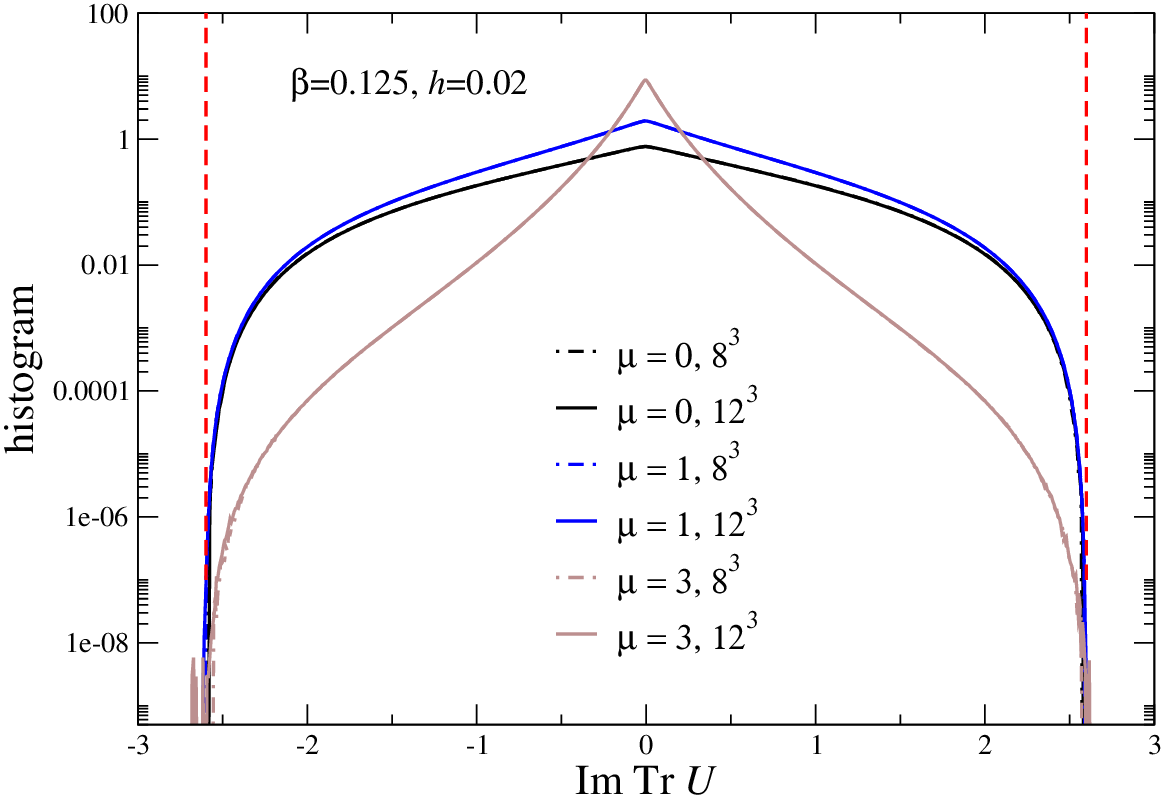} 
 \end{center}
 \caption{Histograms for $\re\Tr U_x$ (left) and $\im\Tr U_x$ (right) at $\beta=0.125$, $h=0.02$ and $\mu=0,1,3$ on $8^3$ and $12^3$.  
 The vertical lines denote the boundaries in SU(3), i.e.\ without complexification. Note the vertical logarithmic scale.    
 }
\label{fig:hist1}
 \end{figure}

As a final assessment, we discuss the extent to which the complexified field space is explored. A sufficiently localized distribution in the imaginary field direction is required for the formal justification to hold \cite{arXiv:0912.3360,arXiv:1101.3270}.
In Fig.\ \ref{fig:hist1} we show histograms for $\re\Tr U$ (left) and $\im\Tr U$ (right), obtained by binning the data sampled during the Langevin process, for three different $\mu$ values and two lattice volumes at $\beta=0.125$. 
For real dynamics, i.e.\ when the angles $\phi_{1,2}$ are real and $U\in$ SU(3), $\Tr U$ is complex-valued, taking values in a triangular shape with corners  at $3e^{2q\pi i/3}$ ($q=0,1,2$). The corresponding  boundaries are shown in the figures as vertical dashed lines and the histograms at $\mu=0$ are contained within these boundaries.
After complexification, $\phi_{1,2}$ are no longer proper angles and 
the dynamics takes place in the larger group SL(3,$\mathbb{C}$). At nonzero $\mu$, we observe that the SU(3) boundaries are indeed crossed, as required, but that the distribution appears to remain localized (note the vertical logarithmic scale). 
The tails of the distributions are noisy, as they are visited during the Langevin process very rarely. There is very little volume dependence.
We also note that the histograms at  $\mu=1$ (in the disordered phase) resemble the histograms at $\mu=0$, while at $\mu=3$ (in the 
ordered phase) they are significantly different, which is reflected in the larger expectation value of $\bra\Tr U\ket$.
The histograms for  $\im\Tr U$ are symmetric within numerical uncertainty, since $\bra\Tr U\ket$ is real.

 \begin{figure}[t]
 \begin{center}
  \includegraphics[width=0.7\textwidth]{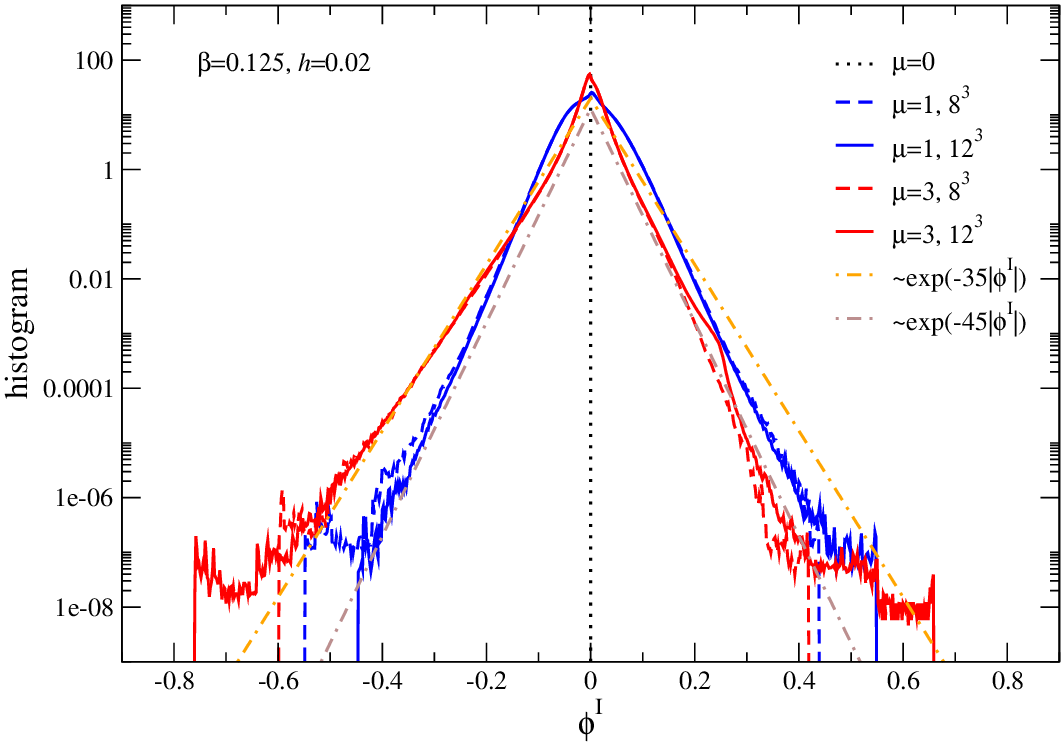} 
 \end{center}
 \caption{Histogram $P(\phi^\rmI)$, where $\phi^\rmI=\{\phi^\rmI_1, \phi^\rmI_2\}$,
  at $\beta=0.125$, $h=0.02$ and $\mu=0,1,3$ on $8^3$ and $12^3$. When $\mu=0$, $\phi^\rmI\equiv 0$.   The dashed straight lines are $P(\phi^\rmI)\sim e^{-b|\phi^\rmI|}$ with $b=35, 45$. Note the vertical logarithmic scale.
 }
\label{fig:hist2}
 \end{figure}
 
Finally, in Fig.\ \ref{fig:hist2} we show the histogram for $\phi^\rmI=\{\phi^\rmI_1, \phi^\rmI_2\}$. For real Langevin dynamics at $\mu=0$, $\phi^\rmI\equiv 0$. At finite $\mu$, nonzero values are generated by the complex drift term. 
We observe that the distribution drops exponentially, over many decades, before the signal becomes noisy.
The straight dashed lines indicate $P(\phi^\rmI)\sim e^{-b|\phi^\rmI|}$ with $b=35, 45$, and are meant to guide the eye. Note that the exponential drop is considerably faster than in the U(1) model studied in Refs.\   \cite{arXiv:0912.3360,arXiv:1101.3270}, where $b\sim 2$ and complex Langevin dynamics failed.
This fast drop and localization of the distribution is another requirement for the applicability of complex Langevin dynamics.\footnote{An open question is what happens to expectation values of the form $\bra\Tr U^k\ket$ with $k$ large.
These observables  contain terms of the form $e^{-k\phi^\rmI}\cos(k\phi^\rmR)$ and the presence of the rapidly oscillating cosine should be taken into consideration, see also Ref.\   \cite{arXiv:1101.3270}.}

%\newpage

\section{Summary and outlook}
\label{sec:sum}

In this paper we revisited the SU(3) spin model, an effective dimensionally reduced Polyakov loop model for QCD in the strong-coupling and heavy-quark limit, at nonzero chemical potential. To handle the sign problem
we employed complex Langevin dynamics, paying special attention to the justification of the method. Using analyticity at small $\mu^2$ (Taylor series expansion and smoothness in $\mu^2$), formal criteria for correctness, and localization of distributions in the complexified space, we arrive at the conclusion that complex Langevin dynamics is reliable in both the ordered and the disordered phase, including the critical region. This should be contrasted with the case of the XY model,
where correct results were obtained in the ordered phase but neither in the disordered phase nor in the transition region \cite{arXiv:1005.3468}.
In the XY model this failure was detected by an apparent lack of analyticity at small $\mu^2$ and the presence of very broad, slowly decaying distributions (as well as by comparing to results obtained in the world line formalism \cite{arXiv:1001.3648}). 
We can therefore conclude that the assessments employed here can be used constructively to rule out or support the applicability of complex Langevin dynamics.\footnote{Nevertheless, it will still be interesting to compare with results obtained in the flux formulation \cite{arXiv:1104.2503,arXiv:1111.0916}.} We emphasize that these tests are generally applicable and not specific to the theory considered here.
 Besides supporting the results of complex Langevin dynamics, we found that the criteria for correctness are also relevant for real Langevin dynamics, as they show clear sensitivity to finite-stepsize effects.
 In order to eliminate the leading-order stepsize dependence, we have successfully implemented a simple higher-order
 algorithm and found it to remove essentially all stepsize dependence in the observables.

How can the different behaviour of the XY model and the SU(3) model under complex Langevin evolution be understood? 
One of the features distinguishing the two is the presence of a non-trivial Haar measure in the SU(3) case. Preliminary results indicate  that it is this measure which leads to more controlled complex  Langevin dynamics. We hope to come back to this in the near future \cite{all}.

%***************************************************************

 \vspace*{0.5cm}
 \noindent
 {\bf Acknowledgments.} \\
We thank Christof Gattringer, Simon Hands, Erhard Seiler, Denes Sexty and Ion-Olimpiu Stamatescu for discussion. This work is carried as part of the UKQCD collaboration and the DiRAC Facility jointly funded by STFC, the Large Facilities Capital Fund of BIS and Swansea University. The authors are supported by STFC.

\appendix

\section{Higher-order algorithm}
\label{sec:appA}

In this appendix, we discuss the higher-order algorithm (\ref{eq:CCC}) in  some more detail. In Ref.\ \cite{CCC} the algorithm is constructed by considering a single update step.  It was also shown that the set  (\ref{eq:CCC}) is part of a more general update rule. 
In order to complement the analysis of Ref.\ \cite{CCC}, we demonstrate the algorithm here using a simple linear kernel. 
We emphasize that the analysis in Ref. \cite{CCC} is for general nonlinear drift term.
For notational simplicity, we use a single degree of freedom $\phi$.

The goal is to solve to higher accuracy
\be
\phi_{n+1} = \phi_n +\eps K(\phi_n) +\sqrt\eps\,\eta_n, \quad\quad\quad \bra\eta_n\eta_{n'}\ket=2\delta_{nn'}.
\ee
Consider the linear kernel $K=-\om\phi$. The solution of the stochastic process, with vanishing initial conditions, is then given by 
\be
\phi_n = \sqrt\eps\sum_{i=0}^{n-1}\left(1-\eps \om\right)^{n-1-i}\eta_i,
\ee
leading to 
\be
\lim_{n\to\infty} \bra\phi_n\phi_n\ket 
 = \frac{1}{\om} \frac{\bra\eta_i\eta_i\ket}{2-\eps \om} 
 =  \frac{1}{\om} \left(1+\half\eps \om+\dots\right),
\label{eq:lo}
\ee
indicating the linear stepsize dependence (we always assume $\eps \om<1$).

Ref.\ \cite{CCC} proposes the following update
\begin{align}
\nn
\psi_n  		=\; & \phi_n +\half\eps K(\phi_n) + k\sqrt \eps\,\tilde\alpha_n, \\
\nn
\tilde \psi_n 	=\; & \phi_n +\half\eps K(\phi_n) + l\sqrt{\eps}\,\tilde\alpha_n, \\
\phi_{n+1} 	=\; & \phi_n+ \eps \left[ a K(\psi_n) + b K(\tilde \psi_n)\right] +\sqrt\eps\, \alpha_n,
\label{eq:CCC2}
\end{align}
where the coefficients  $a$, $b$, $k$ and $l$  are  to be determined, and the noise satisfies
\be
 \tilde\alpha_n = \half\alpha_n+\frac{\sqrt 3}{6}\xi_n,
\ee
and
\be
\bra\alpha_n\alpha_{n'}\ket = \bra\xi_n\xi_{n'}\ket = 2\delta_{nn'}, 
 \quad\quad\quad
\bra\alpha_n\xi_{n'}\ket = \bra\alpha_n\ket =  \bra\xi_n\ket = 0. 
\ee
Note that
\be
\bra\tilde\alpha_n\tilde\alpha_{n'} \ket = \frac{2}{3}\delta_{nn'}, 
\;\;\;\;\;\;\;\;
\bra\alpha_n\tilde\alpha_{n'} \ket = \delta_{nn'}.
\ee
Straightforward substitution in the case of the linear kernel gives
\be
\phi_{n+1} = \phi_n -\eps \tilde \om\phi_n + \sqrt\eps \,\tilde \eta_n,
\ee
with 
\be
\tilde \om = \om(a+b)\left(1-\half\eps \om\right), \;\;\;\;\;\;\;\; \tilde \eta_n = \alpha_n-(ak+bl)\eps \om \tilde\alpha_n.
\ee
Noting that 
\be
\bra\tilde\eta_n\tilde\eta_{n'}\ket = 2\left( 1 - (ak+bl) \eps \om + \frac{1}{3}(ak+bl)^2(\eps \om)^2\right) \delta_{nn'},
\ee
we find, see Eq.\ (\ref{eq:lo}),
\be
\lim_{n\to\infty} \bra\phi_n\phi_n\ket 
 = \frac{1}{\tilde \om} \frac{\bra\tilde\eta_i\tilde\eta_i\ket}{2-\eps \tilde \om} 
 = \frac{1}{\om}\left( \frac{1}{a+b} + \frac{1+a+b-2(ak+bl)}{2(a+b)}\eps \om +\ldots\right).
\ee
We can now determine the coefficients and take
\be
a+b=1, \quad\quad\quad ak+bl=1.
\ee
The resulting expectation value is 
\be
\lim_{n\to\infty} \bra\phi_n\phi_n\ket 
 = \frac{1}{\om}\left( 1- \frac{\eps^2\om^2}{6} +\ldots\right),
\ee
with a remaining correction of  ${\cal O}(\eps^2)$.
The two conditions above do not fully determine the coefficients $a$, $b$, $k$, $l$. Ref.\ \cite{CCC} argues that a further condition
\be
 ak^2+bl^2=\frac{3}{2},
\ee
follows from minimizing the local truncation error, which is beyond the scope of the linear example discussed here. 
In the main part of the paper, the coefficients are taken as
\be
a= \frac{1}{3}, \quad\quad b=\frac{2}{3}, \quad\quad  k=0, \quad\quad l=\frac{3}{2},
\ee
which satisfies the constraints above.

\section{Tables}
\label{sec:tables}

\begin{table}[h]
\begin{center}
\begin{tabular}{ l | l | l l l}
\hline\hline
 & $\eps$ & \;\, $\bra\Tr U\ket$ & \;\, $\bra\Tr U^{-1}\ket$ & \;\, $\bra n \ket$ \\
  \hline 
      		& 0.001     & \;\, 0.2288(22) & \;\, 0.3479(17) & \;\, 0.00988(11) \\
lowest 	& 0.00075 & \;\, 0.2290(23) & \;\, 0.3488(18) & \;\, 0.00988(11) \\ 
order	& 0.0005    & \;\, 0.2369(23) & \;\, 0.3553(18) & \;\, 0.01026(11)  \\
algorithm	& 0.00025 & \;\, 0.2429(27) & \;\, 0.3606(21) & \;\, 0.01055(13) \\
		& 0.0001   & \;\, 0.2396(26) & \;\, 0.3584(20) & \;\, 0.01039(12)   \\ 
                  	& 0.00005 & \;\, 0.2413(21) & \;\, 0.3600(16) & \;\, 0.01047(10)   \\ 
  \cline{2-5}
          & extrapolation & \;\, 0.2419(19) & \;\, 0.3605(15) & \;\, 0.010497(92) \\
 \hline
 		& 0.001      & \;\, 0.2410(27) & \;\, 0.3593(21) & \;\, 0.01046(13) \\
improved  & 0.00075 & \;\, 0.2391(20) & \;\, 0.3579(15) & \;\, 0.01037(10) \\
algorithm	& 0.0005    & \;\, 0.2426(25) & \;\, 0.3606(19) & \;\, 0.01053(12) \\
		& 0.00025	 &  \;\, 0.2415(34) & \;\, 0.3604(26) & \;\, 0.01048(17) \\
                   & 0.0001   & \;\, 0.2407(23) & \;\, 0.3593(18)  & \;\, 0.01044(11) \\                   
  \cline{2-5}
		& average & \;\, 0.2407(11) & \;\, 0.35921(84) & \;\, 0.010443(53)  \\
\hline\hline
& $\eps$& \;\, $\bra L\Tr U\ket$ & \;\, $\bra L\Tr U^{-1}\ket$ & \;\, $\bra Ln\ket$ \\
\hline
     		& 0.001     & \;\, 0.00381(71) & \;\, 0.00644(70)  & \;\, 0.000159(33) \\
 lowest 	& 0.00075 & \;\, 0.00351(67) & \;\, 0.00528(66) & \;\, 0.000152(32) \\
order	& 0.0005   & \;\, 0.00156(69) & \;\, 0.00270(68) & \;\, 0.000065(33)  \\
algorithm	& 0.00025 & $-$0.00029(80)& \;\, 0.00027(78) & $-$0.000018(38) \\
		& 0.0001   & \;\, 0.00094(80) & \;\, 0.00109(79) & \;\, 0.000043(38) \\
                  & 0.00005 & \;\, 0.00056(58) & \;\, 0.00068(57) & \;\, 0.000026(27)   \\
\hline
 		& 0.001      & \;\, 0.00008(74) & \;\, 0.00001(72) & $-$0.000014(35) \\
improved 	& 0.00075 & \;\, 0.00079(60) & \;\, 0.00049(59)  & \;\,  0.000022(28) \\
algorithm	& 0.0005  & $-$0.00017(74) & $-$0.00036(71) & $-$0.000024(35) \\
		& 0.00025 & \;\, 0.0004(12)    & \;\, 0.0001(12)   & \;\,   0.000001(58) \\
                  	& 0.0001  & \;\,  0.00053(69)  & \;\, 0.00026(68)  & \;\,  0.000010(33) \\
\end{tabular}
 \caption{Stepsize dependence for $\mu=1$ (disordered phase), $\beta=0.125$ and $h=0.02$ on a lattice of size $10^3$. 
  }
\label{tab:hpmu1}
\end{center}
\end{table}

In this Appendix we list the results for the stepsize dependence obtained at $\beta=0.125$ and $h=0.02$ on a $10^3$ lattice, for $\mu=1$ (Table \ref{tab:hpmu1}) and $\mu=3$ (Table \ref{tab:hpmu3}). The total Langevin time is 10,000 for $\mu=1$ and 5,000 for $\mu=3$, after discarding the thermalization stage.

Every table shows the real part of the three observables $\bra \Tr U_x\ket$, $\bra \Tr U_x^{-1}\ket$ and $\bra n\ket$ in the upper part, and the real part of the criteria 
$\bra  L \Tr U_x\ket$, $\bra  L  \Tr U_x^{-1}\ket$ and $\bra L  n\ket$ in the lower part, for both the ``lowest-order'' and the``improved'' algorithm. Here lowest-order algorithm refers to the standard discretization (\ref{eq:sslo}), which has corrections that are linear in the stepsize; improved algorithm refers to the higher-order algorithm (\ref{eq:CCC}) of Ref.\ \cite{CCC}. In the case of the lowest-order algorithm we performed a linear extrapolation, using the values at the four smallest stepsizes. Since there is very little stepsize dependence left in the case of the improved algorithm, the average is shown.

\begin{table}[h]
\begin{center}
\begin{tabular}{l |  l | l l l}
\hline\hline
 & $\eps$ & $\bra\Tr U\ket$ & $\bra\Tr U^{-1}\ket$ & $\bra n \ket$ \\
\hline
	       	& 0.001      &  1.69646(39) & 1.73658(35)  & 0.67976(16)  \\
lowest	& 0.00075 &  1.69872(35) & 1.73883(31)  & 0.68066(14)  \\
order 	& 0.0005    &  1.70165(40) & 1.74158(35)  & 0.68184(16)  \\  
algorithm	& 0.00025 &  1.70475(46) & 1.74450(41)  & 0.68308(18)  \\
		& 0.0001   &  1.70442(36) & 1.74427(32)  & 0.68295(15)  \\
                   & 0.00005 &  1.70586(42) & 1.74561(37)  & 0.68352(17)  \\
  \cline{2-5}
	& extrapolation & 1.70615(27)  &1.74590(24)  & 0.68364(11)  \\
\hline
 		& 0.001    & 1.70605(40)  & 1.74571(36)  &  0.68360(16) \\
improved	& 0.00075	& 1.70514(27) & 1.74486(24)  &  0.68324(11) \\
algorithm	& 0.0005  & 1.70629(43) & 1.74598(38)  &  0.68370(17)	 \\	
		& 0.00025	& 1.70597(29) & 1.74571(26)  & 0.68357(12) \\
                  & 0.0001  & 1.70596(39)  & 1.74576(35)  & 0.68356(16) \\
  \cline{2-5}
		& average	 & 1.70576(15)  & 1.74549(13)  & 0.683485(60)   \\
\hline\hline
& $\eps$& $\bra L\Tr U\ket$ & $\bra L\Tr U^{-1}\ket$ & $\bra Ln\ket$ \\
\hline
      		& 0.001      &  0.0480(10)    & 0.05258(97)  &  0.01922(40) \\
lowest 	& 0.00075 &  0.03198(86)  & 0.03488(85)  &  0.01281(34) \\
order 	& 0.0005   &  0.01944(98)  &  0.02119(96)  &  0.00779(39)  \\
algorithm	& 0.00025	 &  0.0084(14)    & 0.0092(13)     & 0.00335(55) \\
		& 0.0001   &  0.0055(11)    & 0.0058(10)     & 0.00219(42)  \\
                  	& 0.00005 &  0.0020(12)     & 0.0021(12)     & 0.00081(50)  \\
\hline
		& 0.001      & 0.0070(11)    & 0.0056(11)   & 0.00189(43) \\
improved	& 0.00075 & 0.00624(88) & 0.00443(85) & 0.00169(35) \\
algorithm	& 0.0005   & 0.0044(12)    & 0.0026(12)   & 0.00096(48)  \\
		& 0.00025 & 0.00285(67) & 0.00100(65) & 0.00040(27) \\
                  & 0.0001    & 0.0020(10)    & 0.0001(10)   & 0.00006(42) \\
\end{tabular}
 \caption{As in Table \ref{tab:hpmu1},  for $\mu=3$ (ordered phase).
  }
\label{tab:hpmu3}
\end{center}
\end{table}

\end{document}